\documentclass[amsmath,amssymb,superscriptaddress]{revtex4-2}

\draft 

\usepackage[pdftex]{graphicx}

\usepackage{xcolor}
\usepackage[utf8]{inputenc}
\usepackage[T1]{fontenc}
\usepackage{mathptmx}
\usepackage{etoolbox}
\usepackage{wasysym}
\usepackage{siunitx}

\usepackage{transparent}
\usepackage[colorlinks=true, citecolor=blue, linkcolor=blue, urlcolor=blue]{hyperref}

\begin{document}


\title[Ultrastable, high-repetition-rate attosecond beamline for time-resolved XUV-IR coincidence spectroscopy]{Ultrastable, high-repetition-rate attosecond beamline for time-resolved XUV-IR coincidence spectroscopy}
\author{D. Ertel}
\email[Corresponding author: ]{dominik.ertel@physik.uni-freiburg.de.}
\affiliation{Institute of Physics, University of Freiburg, 79104 Freiburg, Germany.}
 
\author{M. Schmoll}
\author{S. Kellerer} 
\author{A. Jäger}
\affiliation{Institute of Physics, University of Freiburg, 79104 Freiburg, Germany.}

\author{R. Weissenbilder}
\affiliation{Department of Physics, Lund University, SE-221 00 Lund, Sweden.}

\author{M. Moioli}
\author{H. Ahmadi}
\affiliation{Institute of Physics, University of Freiburg, 79104 Freiburg, Germany.}

\author{D. Busto}
\affiliation{Institute of Physics, University of Freiburg, 79104 Freiburg, Germany.}
\affiliation{Department of Physics, Lund University, SE-221 00 Lund, Sweden.}

\author{I. Makos}
\affiliation{Institute of Physics, University of Freiburg, 79104 Freiburg, Germany.}

\author{F. Frassetto}
\affiliation{CNR-Institute of Photonics and Nanotechnologies (CNR-IFN), 35131 Padova, Italy.}

\author{L. Poletto}
\affiliation{CNR-Institute of Photonics and Nanotechnologies (CNR-IFN), 35131 Padova, Italy.}

\author{C.D. Schröter}
\author{T. Pfeifer}
\author{R. Moshammer}
\affiliation{Max-Planck-Institute for Nuclear Physics, 67119 Heidelberg, Germany.}


\author{G. Sansone}
\affiliation{Institute of Physics, University of Freiburg, 79104 Freiburg, Germany.}


\begin{abstract}
\section*{Abstract}
The implementation of attosecond photoelectron-photoion coincidence spectroscopy for the investigation of atomic and molecular dynamics calls for a high-repetition-rate driving source combined with experimental setups characterized by excellent stability for data acquisition over time intervals ranging from a few hours up to a few days. This requirement is crucial for the investigation of processes characterized by low cross sections and for the characterization of fully differential photoelectron(s) and photoion(s) angular and energy distributions. We demonstrate that the implementation of industrial-grade lasers, combined with a careful design of the delay line implemented in the pump-probe setup, allows one to reach ultrastable experimental conditions leading to an error in the estimation of the time delays of only 12\,as. 
This result opens new possibilities for the investigation of attosecond dynamics in simple quantum systems.

\end{abstract}

\maketitle

\section{Introduction}
In the last years, the development of extreme ultraviolet (XUV) attosecond sources based on high-order harmonic generation (HHG) in gases operating at high repetition rates, ranging from a few tens of kHz up to a few tens of MHz \cite{JPP-Furch-2022, IEEE-Sansone-2012}, has been driven by several applications, including attosecond photoelectron spectroscopy of surfaces \cite{RSI-Wagerl-2011} and photoion-photoelectron coincidence spectroscopy in atoms and molecules \cite{Sabbar2014a, Srinivas2022}. For the first one, space charge effects limit the number of photoelectrons that can be created for each laser shot, while for the second one, the possibility to unambiguously assign to a common molecule or atom the photoelectron(s) and photoion(s) detected, limits the number of photoionization events to (at most) one per laser shot. In both cases, the necessity to reduce the number of charged particles created per laser shot needs to be combined with the need for sufficient data statistics, thus calling for the development of high-repetition-rate infrared (IR) lasers as driving sources for the HHG process.

Due to the low quantum defect and the possibility to use laser diodes for optical pumping, ytterbium (Yb)-based lasers have become in the last years the preferred solution as a driving source for high repetition rate attosecond sources. Even though these systems are characterized by long pulse durations (from a few hundred of femtoseconds to a few picoseconds), the implementation of hollow-core fiber compressors \cite{JOSAB-Beetar-2019}, multi-plate media \cite{APL-Beetar-2018} or multi-pass cells \cite{OPTICA-Viotti-2022} has led to the demonstration of ultrashort and even few-cycle pulses with average powers of few hundreds of Watts \cite{OL-Muller-2021}.

An alternative approach for the generation of attosecond pulses at high-repetition rates is based on optical parametric amplifiers \cite{RSI-Cerullo-2003}. Using a carrier-envelope-phase stabilized laser seed, the generation of few-cycle pulses at hundreds of kHz \cite{OL-Furch-2017, OE-Prinz-2015, NP-Mikaelsson-2021} and even MHz repetition rate \cite{OE-Rothhardt-2012} was demonstrated. These driving pulses can be used for the generation and characterization of isolated attosecond pulses at unprecedented repetition rate \cite{OPTICA-Witting-2022}.

Photolectron-photoion coincidence spectroscopy, based on reaction microscopes (ReMis) \cite{Ullrich1997,Ullrich2003} or cold-target recoil-ion momentum spectroscopy (COLTRIMS) setups \cite{Dorner2000}, is a powerful technique for the investigation of relaxation mechanisms in dimers \cite{PRL-Jahnke-2004}, for the characterization of electronic correlation in double photoionization processes \cite{PRL-Staudte-2007,PRL-Rudenko-2007}, and the imaging of molecular dynamics with a few femtosecond temporal resolution \cite{SCIENCE-Wolter-2016}. When combined with attosecond pulses, photolectron-photoion coincidence spectroscopy has enabled the characterization of attosecond time delays in photoionization in the recoil frame \cite{Heck2021, Ahmadi2022}, the time-resolved investigation of molecular chirality \cite{JPB-Comby-2020}, and the characterization of attosecond photoionization dynamics in size-selected clusters \cite{NATURE-Gong-2022}.

In this work, we demonstrate a novel high-repetition rate attosecond source 
optimized for XUV-IR coincidence spectroscopy. The main advantage of our setup is the excellent stability of the experimental conditions, due to the implementation of an industrial-grade Yb driving source and of a collinear, monolithic delay line, which eliminates short-term instabilities and long-term drift of the relative delay in the pump-probe interferometer. These characteristics allow the continuous acquisition of experimental data for several hours without the need for any active stabilization system.


\section{Experimental Setup} \label{sec:Setup}

    \subsection{Femtosecond light source}
\begin{figure}[htb]
            \centering
            \includegraphics[scale=1]{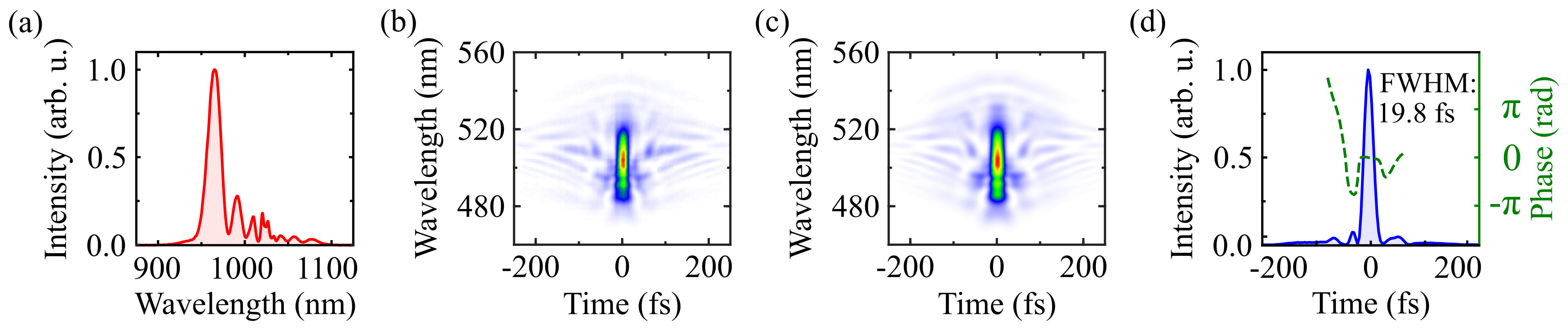}
            \caption{(a) Measured spectrum of the broadened IR pulses using about 3\,bar of Kr in the HCF. (b) Measured and (c) reconstructed FROG trace (FROG error $=0.56$\,\%) after spectral broadening and temporal compression of the IR pulses. (d) Retrieved temporal intensity profile (blue solid line) and phase (green dashed line) of the compressed pulses exhibiting a duration of 19.8\,fs (at FWHM).}
            \label{fig:HCF_Output}
        \end{figure}

        The driving light source of the attosecond beamline is a commercially available Yb:KGW regenerative amplifier ($\lambda_c=\SI{1025}{\nano\meter}$, Pharos PH1-20, Light Conversion Ltd.) producing 277\,fs short laser pulses with an energy of up to \SI{400}{\micro \joule} at 50\,kHz repetition rate. Such pulses are focused using a plano-convex lens ($f=\SI{1}{\meter}$) into a krypton-filled hollow-core fiber \cite{Nisoli1996} ($l=\SI{1}{\meter}$, $d_\text{core}=\SI{250}{\micro\meter}$) for nonlinear spectral broadening. The spectrum, broadened using about 3\,bar of krypton in the HCF, is measured with a commercial spectrometer (SD2000, Ocean Optics), as shown in Fig.~\ref{fig:HCF_Output} (a). After the fiber, a concave mirror ($f=\SI{0.8}{\meter}$) collimates the beam and the pulses are temporally compressed by a set of eight chirped mirrors (PC1611, Ultrafast Innovations) providing a group delay dispersion of around \SI{-150}{\femto\second\squared} per reflection.
        %
        Due to losses during the nonlinear self-phase modulation process, coupling, and transmission through the HCF, as well as the reflection of the chirped mirrors, the pulse energy is reduced to about 230\,\textmu J resulting in a total throughput of 58\,\%.

        Temporal characterization of the IR pulses is performed using a home-built frequency-resolved optical gating (FROG) \cite{Trebino1997} device based on second-harmonic generation. The measured and reconstructed FROG traces are shown in Fig.~\ref{fig:HCF_Output} (b) and (c), respectively. The relatively small FROG error of about 0.56\,\% ensures the reliability of the reconstruction. Additionally, the retrieved temporal profile of the compressed IR pulse and its phase are presented in Fig.~\ref{fig:HCF_Output} (d).
        A pulse duration below \SI{20}{\femto\second} is retrieved from the FROG algorithm as shown in Fig.~\ref{fig:HCF_Output} (c).
        
        In general, by controlling the gas pressure inside the HCF and the number of chirped mirrors used, we can tune the spectral broadening and hence the pulse duration required in the experiment.
        
        %

    \subsection{Attosecond beamline}
\begin{figure*}[t]
            \centering
            \includegraphics[scale=1]{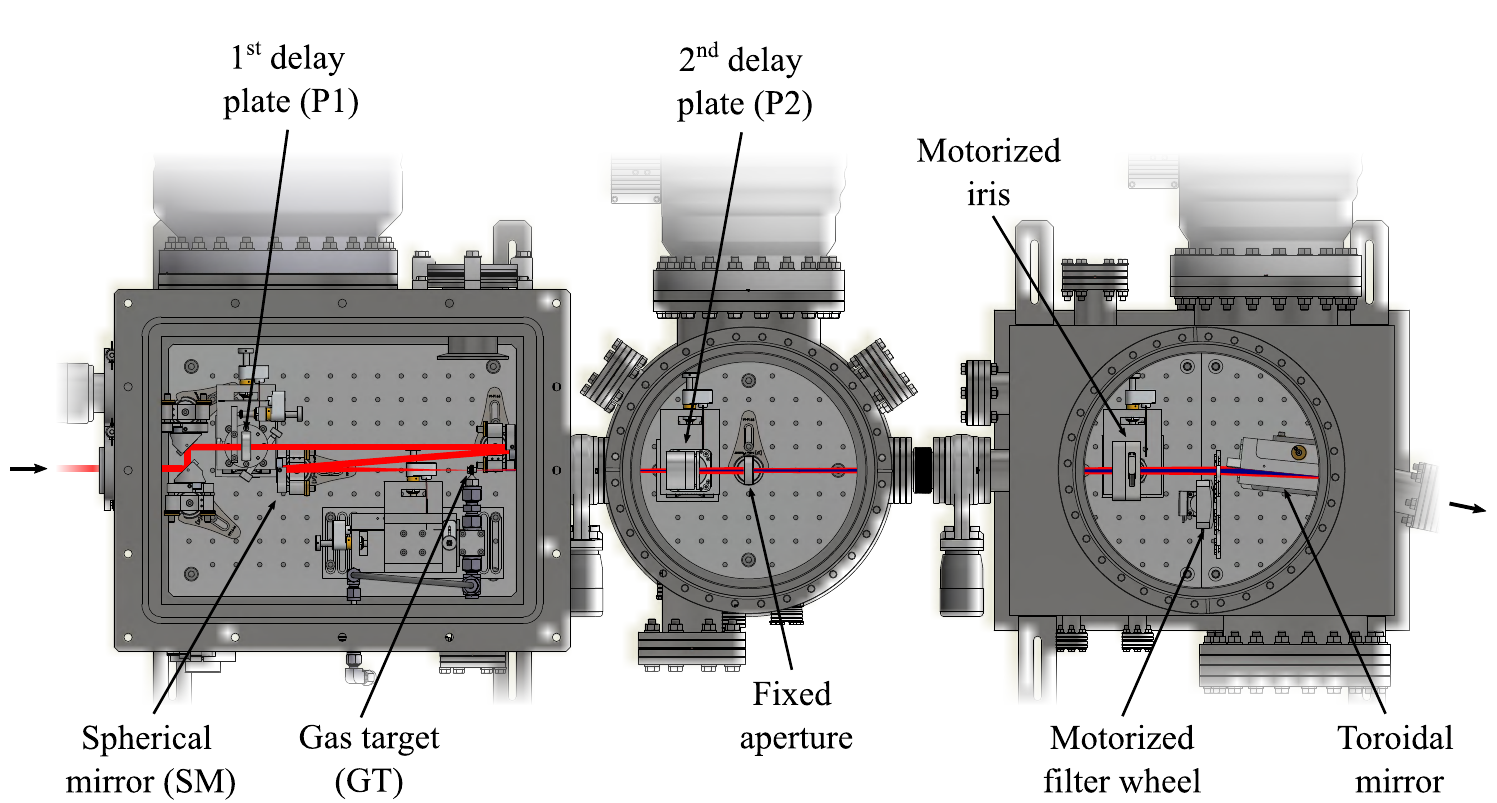}
            \caption{Schematic of the attosecond beamline for coincidence spectroscopy including the incoming IR (red) and generated XUV (blue).}
            \label{fig:beamline}
        \end{figure*}

        The compressed IR pulses are then sent to our attosecond beamline consisting of three vacuum chambers schematically depicted in Fig.~\ref{fig:beamline}. 
        Pinholes acting as differential pumping stages between the individual chambers are installed, ensuring the ultra-high vacuum conditions required for coincidence measurements (typically on the order of $10^{-9}$\,mbar in the ReMi placed after the toroidal mirror chamber at full gas load from the HHG target).
        
        The spectrally broadened and temporally compressed laser pulses enter the first vacuum chamber through a 1-mm-thick fused silica window.
        After the first drilled delay plate P1 (described in Sec.~\ref{sec:delay_line}), the beam is focused into the gas target using a spherical mirror ($f=\SI{20}{\centi\meter}$). 
        The beam is impinging the mirror with an angle of incidence below 3° in order to minimize astigmatism.
        The position of the targets can be adjusted by using motorized translation stages (UMR5.16 and 8301-V, Newport) in all three dimensions. A motorized iris, placed before the beamline (not shown), is used to adjust the IR intensity in the generation medium. The generated XUV and the remaining IR 
        co-propagate 
        through the second drilled delay plate P2 (see Sec.~\ref{sec:delay_line}). Another adjustable motorized iris in the third chamber is used to control the intensity of the IR probe pulse. Using ultra-thin ($\approx 100$\,nm) metallic filters mounted in a motorized filter wheel, we can either fully or partially block the remaining IR, spatially define the XUV beam, and partially compensate for the attochirp \cite{Lopez-Martens2005}. A gold-coated toroidal mirror ($f=\SI{45}{\centi\meter}$) used at a grazing incidence of 84° refocuses the XUV and the eventual IR beam in a $2f-2f$ configuration into the gas target of the ReMi (described in Sec.~\ref{sec:ReMi}).

    \subsection{Compact, flat-field XUV spectrometer}
        Spectral characterization of the generated XUV radiation is performed using a home-built XUV spectrometer installed behind the coincidence spectrometer. It consists of a concave flat-field reflection grating (600 grooves/mm, 001-0639, Hitachi) with \SI{46.9}{\centi\meter} focal length, a microchannel plate (MCP) detector (F2225-11P563, Hamamatsu) with attached phosphor screen and a CMOS camera (C11440-36U ORCA-spark, Hamamatsu). At grazing incidence of \SI{85.3}{\degree}, the incoming XUV is spectrally dispersed and additionally focused in one dimension by the gold-coated grating. Some of the harmonics reach the MCP detector and phosphor screen, which is imaged by the camera. By translating the detector assembly along the focal plane of the diffraction grating, the whole spectral range covered by the XUV beam can be measured.

    \subsection{Ultrastable, collinear XUV-IR delay control}\label{sec:delay_line}
\begin{figure}[htb]
            \centering
            \def\svgwidth{0.9\textwidth} 
            \includegraphics[scale=1]{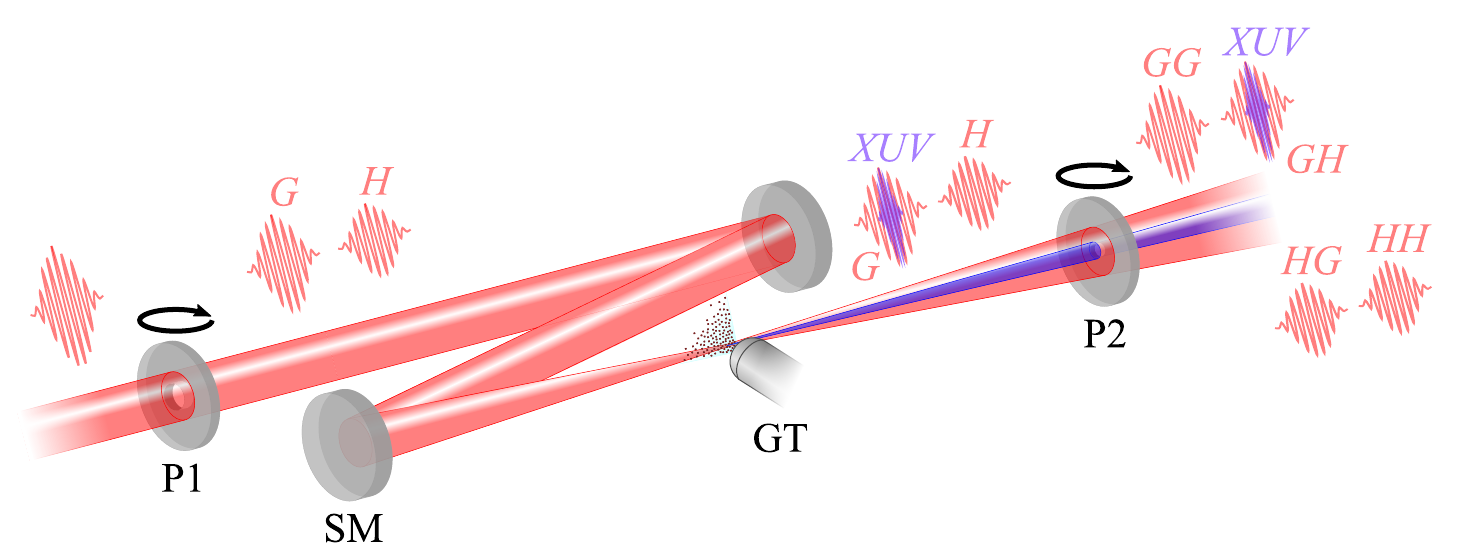}
            \caption{Visualization of the beam propagation through the developed HHG setup illustrating the working principle of the ultrastable collinear attosecond XUV-IR delay line. SM: spherical mirror; GT: gas target. A detailed description can be found in the text.}
            \label{fig:delay-line}
        \end{figure}

        In order to control the delay between XUV pump and IR probe, a collinear geometry for the interferometer is chosen offering high delay stability without the need of active stabilization methods. 
        Our developed collinear XUV-IR interferometer depicted in Fig.~\ref{fig:delay-line} consists of two centrally drilled fused silica plates with thickness $d\approx\SI{1}{\milli\meter}$. A detailed description of the working principle, as well as supporting simulations of the beam propagation through the developed delay line, has been discussed by Ahmadi \textit{et al}. \cite{Ahmadi2020}. The first plate (P1), placed before the spherical mirror (SM), 
        splits the incoming IR laser beam into two spatially and temporally separated parts. Labels ($G$) and ($H$) indicate the propagation through the glass or hole, respectively. The hole diameter of P1 ($\diameter_\text{P1} = \SI{2}{\milli\meter}$) was determined in such a way that the central part ($H$) is weak enough at focus not to drive the HHG process, and thus only the annular part ($G$) contributes to it. The generated XUV (shown in purple color) propagates collinearly with the two IR beams towards the second plate (P2) with \SI{1}{\milli\meter} hole diameter placed \SI{20}{\centi\meter} after the gas target. Passing P2 leads to a further splitting of the IR, resulting in four beams ($GG$, $HG$, $GH$, and $HH$) separated in space and time. Because of the lower divergence of the XUV, most of it is transmitted through the hole in P2. However, due to the tight focusing geometry and depending on the phase-matching conditions, the outermost part of the XUV (originated mainly from the long trajectory) is blocked by the glass of plate P2. Only the pulses $HG$ and $GH$ can temporally overlap with the XUV. By tilting P2 using a high-precision motorized rotation stage (SR-4513-S-HV, Smaract), we can accurately control the time delay $\tau$ between the XUV and the IR dressing pulse $HG$ according to \cite{Zair2018, Ahmadi2020}
        \begin{align}
		    \begin{aligned}
		    \tau = \frac{d}{c}\left[\frac{n}{\cos{\beta}}
		    +\left(\tan{\alpha}-\tan{\beta}\right)\cdot\sin{\alpha}-\frac{1}{\cos{\alpha}}\right] -\tau_{\text{P1}}\,.
	    	\end{aligned}
	    	\label{eq:time-delay}
		\end{align}
        Here, $c$ denotes the speed of light, $n$ the refractive index of P2 at \SI{1025}{\nano\meter}, $\alpha$ ($\beta$) the incident (refractive) angle at the P2 interface, and $\tau_{\text{P1}}$ the time delay accumulated during the transmission through P1. The time delay of the $GH$ pulse with respect to the XUV cannot be varied. Nevertheless, the intensity of the $GH$ pulse is orders of magnitude lower than that of $HG$ and thus its experimental contribution is negligible \cite{Ahmadi2020}. The pulse $GG$ ($HH$) is weak and arrives temporally much after (before) the XUV ($\sim\SI{4.88}{\pico\second}$) and consequently, does not affect our measurements. 
        
    \subsection{Coincidence spectrometer}\label{sec:ReMi}
    
        For performing time-resolved XUV-IR coincidence spectroscopy, we combine our attosecond light source with a ReMi \cite{Ullrich2003}, which allows us to characterize the three-dimensional momenta of charged particles originating from photoionization events.
        
        The gas target is formed by the supersonic expansion of an atomic or molecular gas through a nozzle ($\diameter_\text{orifice} =\SI{30}{\micro \meter}$) and two skimmers leading to a well-confined, cold gas jet. The XUV and IR beams cross and interact with this jet resulting in the production of charged particles (electrons and ions) through photoionization. Weak electric and magnetic fields accelerate and guide these particles to the corresponding detectors. By using time- and position-sensitive particle detectors (MCPs plus delay line anode), both the transverse impact position, as well as the arrival time (time-of-flight, TOF) of the charged particles can be detected. This allows us -- under the assumption of homogeneous electric and magnetic fields -- to reconstruct the three-dimensional momenta of such particles using classical calculations for particle motion \cite{Ullrich2003}. 
        
\section{Characterization of the generated high-order harmonics}
        Efficient HHG with driving pulse energies in the range of few hundreds of \si{\micro\joule} requires a relatively tight focusing geometry (here: \SI{20}{\centi\meter} focal length).
        However, such a focusing geometry leads to a relatively small Rayleigh length in the order of 1\,mm implying a short but dense interaction medium.
        Two different types of HHG gas targets are available, a gas nozzle ($\diameter_\text{orifice} = \SI{50}{\micro\meter}$) operating at up to \SI{10}{\bar} backing pressure and a gas cell ($l=\SI{1}{\milli\meter}$) operating at up to \SI{1}{\bar} backing pressure. Both targets provide noble gases (here: argon) as a nonlinear generation medium.
        
\begin{figure*}[t]
            \centering
            \includegraphics[scale=1]{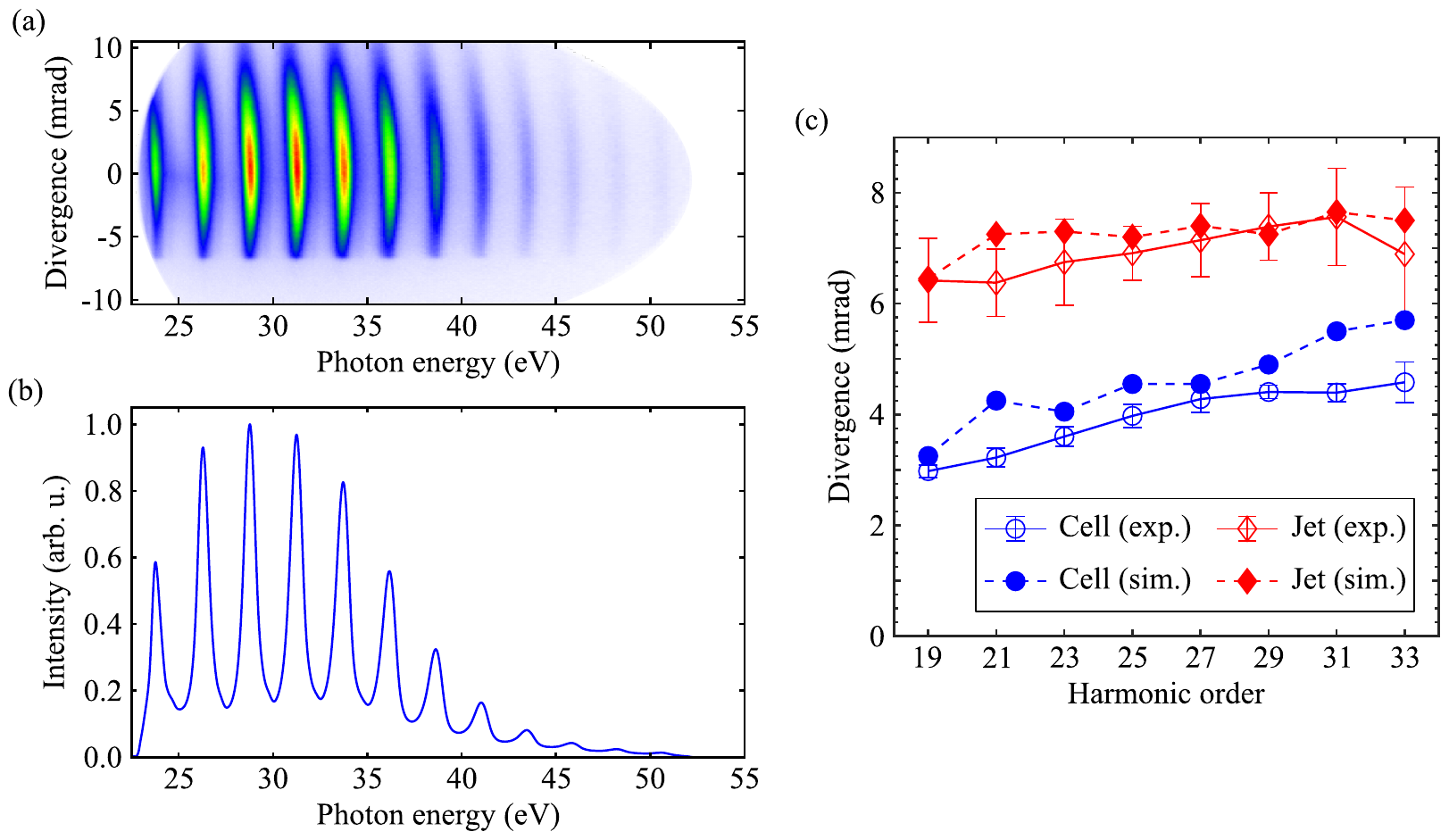}
            \caption{(a) Spectrum of the high-order harmonics generated in argon using a gas nozzle with a 50\,\textmu m orifice and recorded with our home-built XUV spectrometer. The second delay plate P2 was not installed in the beam path. Note that the distortion of the usually circular MCP originates from rescaling the spectrum to an equidistant energy axis. (b) Corresponding section of the XUV spectrum. The fundamental IR beam was blocked by transmission through an 0.2\,\textmu m thick aluminum filter. (c) Divergence of the experimentally generated and theoretically simulated XUV as a function of the harmonic order for both available gas targets.}
            \label{fig:HHG_50mumNozzle}
        \end{figure*}

        As an example, Fig.~\ref{fig:HHG_50mumNozzle} (a) depicts a section of the individual high-order harmonics generated in argon using the gas nozzle and recorded with our home-built XUV spectrometer. The corresponding section of the XUV spectrum is shown in (b). For this measurement, as well as for all results presented in this section, the second delay plate P2 was not installed in the beam path and we blocked the fundamental IR beam using an 0.2\,\textmu m thick aluminum filter.
        
        Considering that the toroidal mirror images with unity magnification the XUV source point and that the flat-field concave grating utilized in our XUV spectrometer only focuses the XUV in one dimension (horizontal one corresponding to the spectral dispersion direction), we can directly relate the spatial expansion in the vertical dimension on the MCP detector to the original divergence of the individual harmonics.
        Figure~\ref{fig:HHG_50mumNozzle} (c) presents the experimental divergence of the generated XUV radiation as a function of the harmonic order for the two gas targets. The data consist of several measurements under the same experimental conditions carried out on different days, from which the mean value along with the associated standard deviation was then determined. Results of numerical simulations of the HHG process performed applying the method described 
        in ref. \cite{Weissenbilder2022a}, 
        using simulation parameters which reproduce the experimental conditions, 
        are additionally depicted in Fig.~\ref{fig:HHG_50mumNozzle} (c). The agreement between experiments and numerical predictions is very good, indicating that the divergence increases with rising harmonic order for both gas targets under the present experimental conditions.
        
        Similar studies \cite{Quintard2019, Wikmark2019, Chatziathanasiou2019, Hoflund2021} 
        investigating the spatial characteristics of high-order harmonic beams have reported the dependence of the XUV divergence (and focusing properties) on the longitudinal position of the generating medium, as well as on the harmonic order. Such dependencies can be ascribed to the wavefront curvature of the IR-generating beam and that induced by the intensity-dependent dipole phase \cite{Wikmark2019}.
        
        Furthermore, Fig.~\ref{fig:HHG_50mumNozzle} (c) shows that, in our experimental configuration, a higher XUV divergence originates from the gas jet compared to those from the gas cell. The reason for this relies on the different positions of the gas targets (with respect to the IR focus) and thus on different phase-matching conditions. Even though both targets were positioned behind the IR laser focus, the gas cell needs to be closer to it than the gas jet in order to achieve comparable XUV intensities as an interplay of optimum phase matching conditions \cite{Makos2020}.
        
        
        The experiment presented in the following Sec.~\ref{sec:He_Experiments} of this work is carried out using high-order harmonics generated by the gas jet. The reason for this relies on the better long-term stability of the XUV intensity compared to the gas cell, especially after clipping the XUV light with the second delay plate P2 and thus discriminating that radiation part produced from the long electron trajectories.

\section{Application in coincidence spectroscopy}\label{sec:He_Experiments}
     
     \subsection{Time-resolved XUV-IR photoionization experiment}
\begin{figure}[htb]
            \centering
            \includegraphics[scale=1]{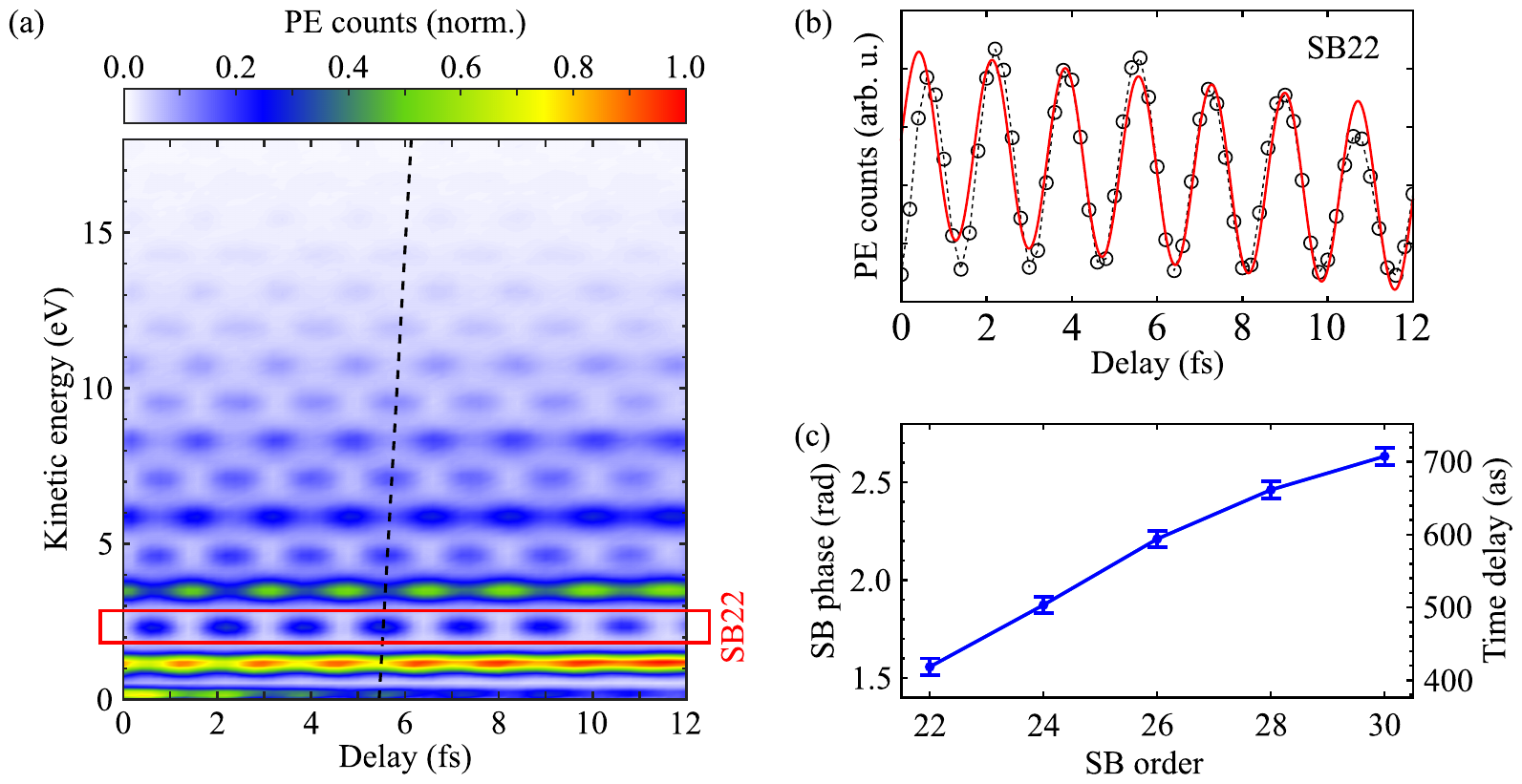}
            \caption{(a) RABBIT trace obtained in helium. The black dashed line with a slope of about 30\,as/eV visualizes the phase shift between successive sidebands (SBs). (b) Delay-dependent oscillation of SB22 (black circles, dashed line) extracted from the RABBIT spectrogram in (a) and corresponding fit (red line) according to Eq.~\ref{eq:SB}. (c) SB phases $\Delta\varphi_\text{SB}$ and associated time delays $\tau_\text{SB}$ retrieved from the fit of the oscillations of the first five SBs visible in the RABBIT trace.}
            \label{fig:He_RABBIT__Phases_combined}
        \end{figure}

        Combining the high-repetition-rate HHG source including the highly stable XUV-IR delay interferometer together with a three-dimensional photoelectron/-ion coincidence spectrometer
        enables to investigate dynamical processes during the photoionization of atoms and molecules on their natural time scale and with full solid angle. In a first proof-of-principle measurement demonstrating the capability of our experimental setup, we performed photoionization studies based on the interferometric RABBIT (reconstruction of attosecond beating by interference of two-photon transitions) technique \cite{Paul2001} in helium. The fundamental principle of such measurement technique relies on two-photon two-color ionization. An XUV photon of the attosecond pulse train (APT) ionizes the sample under investigation; due to the absorption or emission of an additional IR probe photon, the released photoelectron (PE) undergoes a continuum-continuum transition populating the so-called sidebands (SBs). The yield of such SBs depends on the XUV-IR delay $\tau$ according to
            \begin{align}
                I_\text{SB}(\tau) \propto \cos{\left(2\omega_\text{IR}\tau-\Delta\varphi_\text{SB}\right)}\,,
            \label{eq:SB}
            \end{align}
        where $\omega_\text{IR}$ denotes the fundamental IR frequency and $\Delta\varphi_\text{SB}$ the SB phase consisting of the attochirp $\Delta\varphi_\text{XUV}$ and the atomic phase $\Delta\varphi_\text{atomic}$.
        By recording PE spectra as a function of the XUV-IR delay, one acquires a RABBIT spectrogram.
        Figure~\ref{fig:He_RABBIT__Phases_combined} (a) shows a RABBIT spectrogram obtained in helium with an acquisition time of about 6.5 hours. 
        
        
        By integrating the individual SB signals over a certain energy range and subsequently fitting the SB oscillations according to Eq.~\ref{eq:SB} (shown in Fig.~\ref{fig:He_RABBIT__Phases_combined} (b) for SB22), we can access the phase $\Delta\varphi_\text{SB}$ of each SB. The resulting $\Delta\varphi_\text{SB}$ (for the first five SBs) are plotted over the corresponding SB order in Fig.~\ref{fig:He_RABBIT__Phases_combined} (c). Here, the error bars correspond to the standard deviation returned by the fitting algorithm. 
        The linear increase of the SB phases (about 30\,as/eV), highlighted by the black dashed line in Fig.~\ref{fig:He_RABBIT__Phases_combined} (a), 
        originates mainly from the uncompensated attochirp of the attosecond pulse train.
        
        The averaged value of the errors in the retrieved SB phases is in the order of 40\,mrad, corresponding to an error of around 12\,as in the time delay. Such small error indicates a remarkable long-term stability of our system making it an excellent tool for attosecond time-resolved coincidence spectroscopy utilizing the RABBIT technique.
        
    \subsection{Angle-resolved photoionization dynamics in helium}    
        %
        The ReMi offers the possibility to measure the momenta of the charged particles, created during the photoionization process, with full $4\pi$ solid-angle collection. This allows for studying the photoionization dynamics as a function of the emission angle of the released PE \cite{Heuser2016, Heck2021, Ahmadi2022}. In order to demonstrate this capability of our system, we investigated the XUV-IR angle-resolved photoionization process in helium.
        
\begin{figure}[htb]
            \centering
            \includegraphics[scale=1]{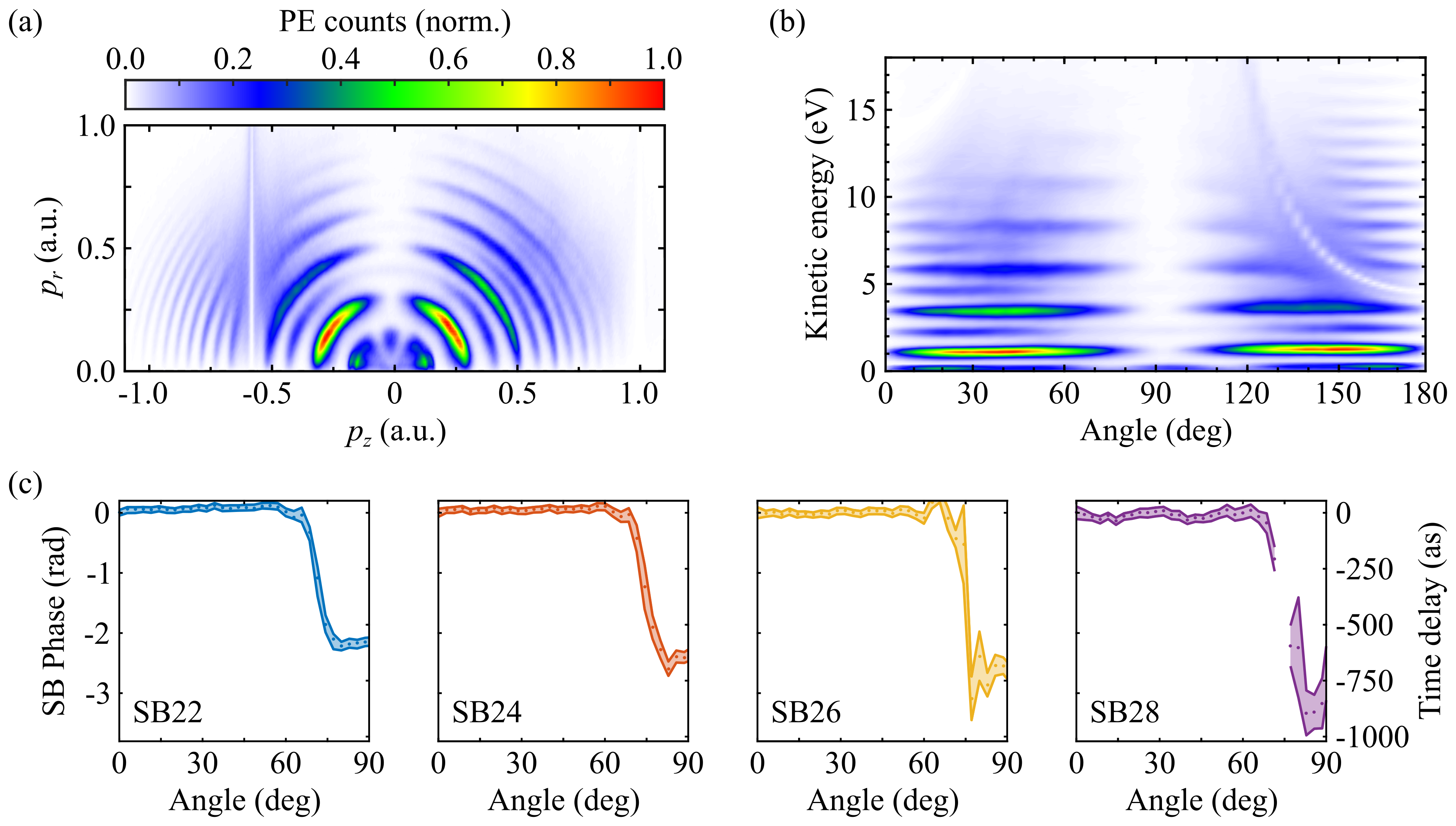}
            \caption{(a) PE momentum distribution, i.e. normalized PE counts as a function of the radial PE momentum $p_r$ and the momentum component $p_z$ along the $z$-direction of the ReMi spectrometer, for helium. (b) Distribution of the PE counts as a function of the PE kinetic energy and PE emission angle. The vertical line close to $p_z=-0.6$\,a.u. in (a) represents a loss in the PE momentum information due to a magnetic node, which is also reflected in (b) between 120° to 180°. The color scale in (a) also applies for (b). (c) SB phases $\Delta\varphi_\text{SB}$ and associated time delays $\tau_\text{SB}$ as a function of the PE emission angle. The colored shaded areas corresponds to the uncertainty of the analytical fit. Here, only data points with an error smaller than 850\,mrad are shown. The y-axis on the left for SB22, as well as that on the right for SB28, applies for all shown SBs.}
            \label{fig:He_angle_resolved_complete}
        \end{figure}

        We present in Fig.~\ref{fig:He_angle_resolved_complete} (a) the normalized PE momentum-resolved distribution and in (b) that of the PE counts as a function of the PE kinetic energy and PE emission angle measured during the photoionization of helium. In the measurements, the XUV and IR pulses are linearly polarized with the common polarization axis oriented along the z-direction. One can clearly distinguish the more intense PE harmonics describing $p$-waves with no signal at $p_z=0$, i.e. at 90° with respect to polarization direction of the two fields from the weaker SBs representing a combination of $s$- and $d$-waves. The sharp vertical line close to $p_z=-0.6$\,a.u. in Fig.~\ref{fig:He_angle_resolved_complete} (a) represents a loss in the PE momentum information,  originating from those PEs performing an exact integer multiple of cycles of their cyclotron motion \cite{Ullrich2003}. This information loss is also reflected in (b) between emission angles of 120° to 180°.
        
        
        Figure~\ref{fig:He_angle_resolved_complete} (c) shows that the retrieved phases of the first four SBs, visible in the angle-integrated RABBIT trace in Fig.~\ref{fig:He_RABBIT__Phases_combined}, varies as a function of the PE emission angle, in agreement with previously reported studies \cite{Heuser2016, Busto2019}. Note that the phases of each SB are subtracted by the corresponding SB phase at 0° emission angle. The phases of all SBs are approximately independent of the emission angle up to 60° after which they experience a sharp jump. Such a significant anisotropy towards $90^\circ$ emission angle originates from different interference of the $s$- and $d$-partial waves in the absorption and emission path as predicted by Fano's propensity rule for continuum-continuum transitions \cite{Busto2019}. The exact position of the phase jump increases with the SB order as a result of the lower asymmetry between absorption and emission for increasing PE kinetic energy. The increase of the magnitude of the phase jump with SB order originates from the reduced angular momentum dependence of the continuum-continuum phase \cite{Fuchs2020, Peschel2022}.
        At larger PE emission angles, the fit of the SB oscillations becomes decreasingly accurate, due to the lower count rate of the SB signals. This observation explains the larger uncertainties towards higher emission angles, at least for higher SB orders, as shown in Fig.~\ref{fig:He_angle_resolved_complete} (c).
        
        
\section{Summary and future perspectives}
    In conclusion, we have presented a high-repetition-rate attosecond beamline combined with a collinear XUV-IR interferometer and a ReMi. The high repetition rate of 50\,kHz, enabled by the Yb-based femtosecond front-end, is beneficial due to improved statistics during acquisition. Furthermore, the compact and collinear design of the interferometer guarantees high stability of the XUV-IR delay over the long acquisition times required by coincidence spectroscopy.
    
    In order to demonstrate the capability of our system, we studied, in a proof-of-principle measurement, the photoionization of helium applying the RABBIT technique. We first investigated the phases of the SB oscillations integrated over all PE emission angles. The accuracy in the phase determination of about 40\,mrad, corresponding to about 12\,as, proves the excellent performance of our system. The possibility of studying the photoionization process in an angle-resolved manner is additionally presented. Here, the observed anisotropic behavior of the photoionization dynamics in helium is consistent with previously reported experimental data, but thanks to the improved statistics we can measure the photoionization delays at all PE emission angles.
    
    Besides photoionization studies of other atomic systems, the system will allow the time- and angle-resolved investigation of photoionization processes of various molecules with increasing complexity in the laboratory, as well as in the molecular or recoil frame.

\begin{acknowledgments}
    This project has received funding from the Deutsche Forschungsgemeinschaft (DFG) (IRTG CoCo (2079), INST 39/1079 (High-Repetition-Rate Attosecond Source for Coincidence Spectroscopy), QUTIF SA 3470/2, RTG 2717 DynCAM). Funding from the Bundesministerium für Bildung und Forschung (Project 05K19VF1), as well as from the Georg H. Endress foundation and the Swedish Research Council (2020-06384), is gratefully acknowledged.
    
    We thank Anne L'Huillier from the department of physics of Lund University in Sweden for multiple discussions regarding the understanding of the difference in harmonic divergence between cell and jet.
    
\end{acknowledgments}



\bibliography{library_final}

\end{document}